\documentclass{aa}
\usepackage[varg]{txfonts}
\bibpunct{(}{)}{;}{a}{}{,} % to follow the A&A style
\begin{document}
\title{The Hipparcos parallax for Polaris}
\author{Floor van Leeuwen}
\institute{Institute of Astronomy, Cambridge}
\date{Received <date> /Accepted <date>}
\abstract{
This letter follows a recent claim that the Hipparcos parallax for Polaris could be too small by 2.5~mas. It examines in detail the Hipparcos epoch astrometric data for Polaris, as well as the viability of other observations that were put forward to support a larger parallax. The Hipparcos determination of the Polaris parallax is shown to be sufficiently robust to fully exclude a significantly larger parallax, and there is no observational support from other observations, such as a supposed presence of a cluster, either.} 
\maketitle
\section{Introduction}
In a recent letter to the Astrophysical Journal, \citet{2013ApJ...762L...8T} (TKUG from here on) suggested that the parallax as measured by Hipparcos \citep{2007A&A...474..653V, 2007ASSL..350.....V} for \object{Polaris} (HIP~11767, HD~8890) is significantly lower than it should be. The distance of $99\pm 2$~pc suggested by TKUG on the basis of the assumed pulsation mode of Polaris is equivalent to a parallax of $10.1\pm 0.2$ mas, very different from the parallax as measured by Hipparcos, $7.54\pm0.11$~mas. Consequently, I have recently frequently been asked if it is at all possible for the Hipparcos parallax measurement to be so far off. This letter shows the Hipparcos astrometric solution for Polaris in all detail as a means to assess the robustness of that solution, to assess whether its measurement of the parallax could be offset by 23 times its standard error. It also briefly discusses other arguments that have been used to suggest a significantly shorter distance for Polaris than what has been measured by Hipparcos.
\section{Hipparcos epoch astrometry for Polaris}
\begin{figure}[h]
\centering
\includegraphics[width=8cm]{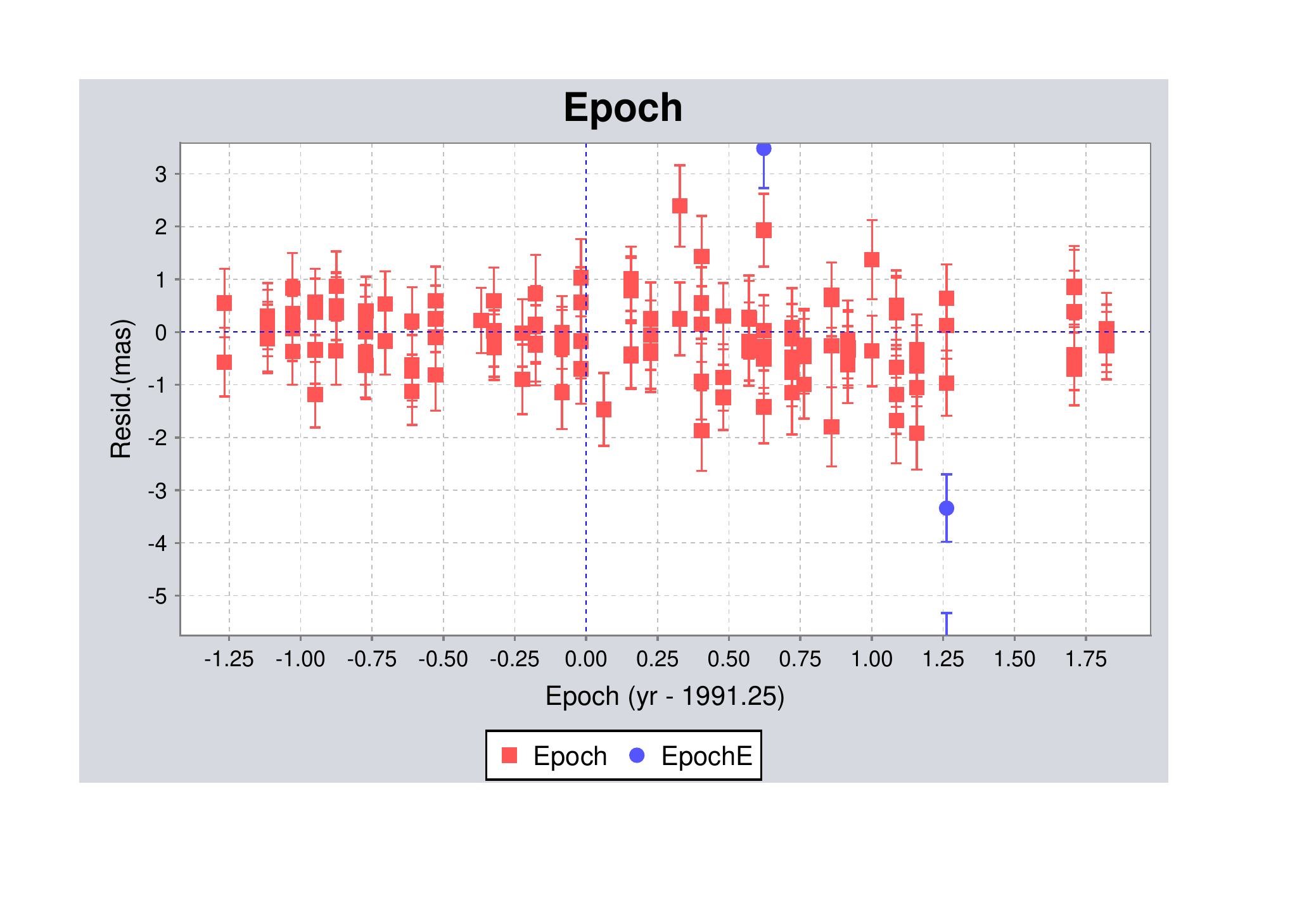}
\caption{Distribution over epoch of the astrometric solution residuals of the Hipparcos measurements for Polaris. Blue dots represent rejected measurements.}
\label{fig:epochdistr}
\end{figure}

\begin{figure}[h]
\centering
\includegraphics[width=6.5cm]{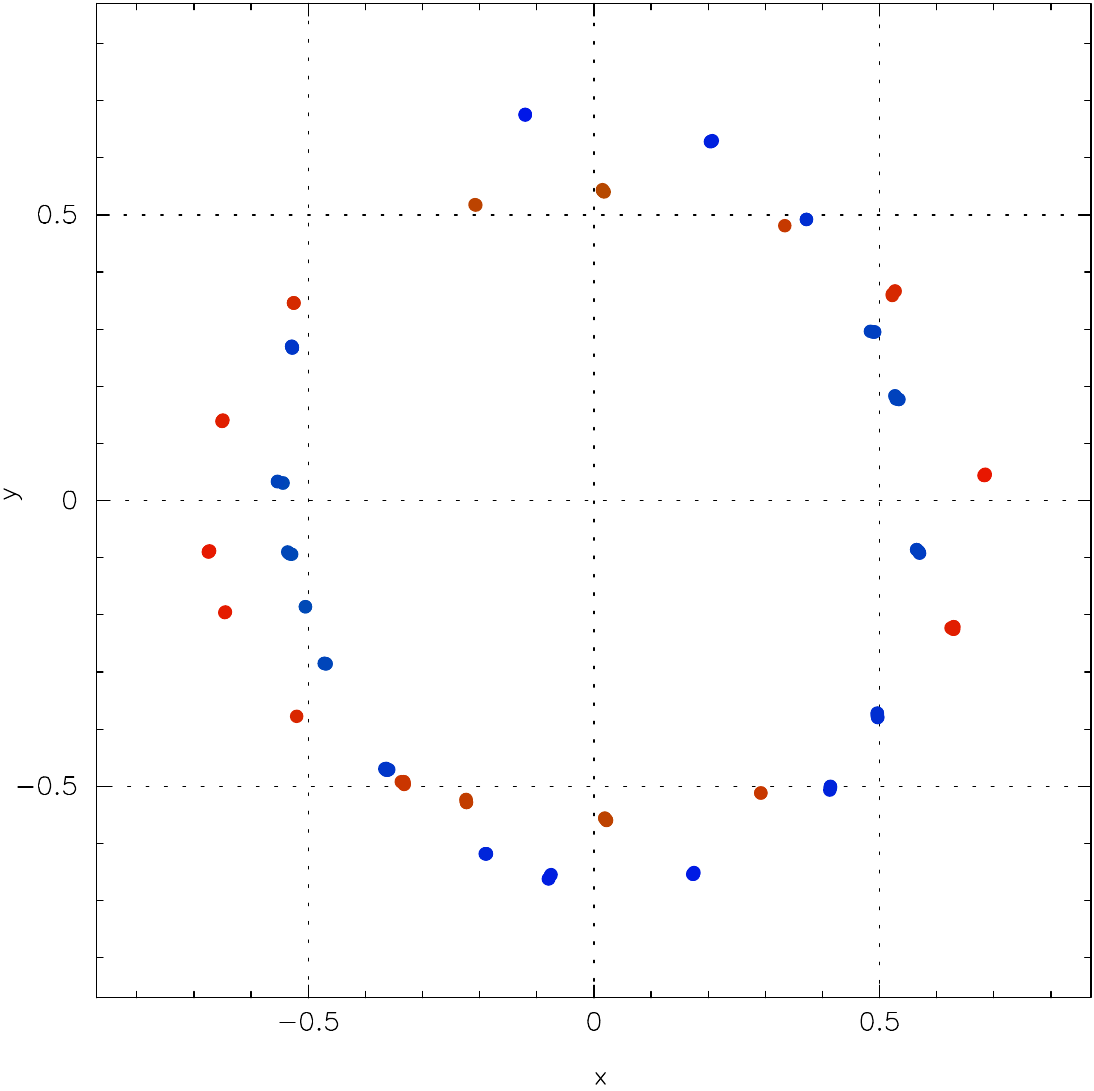}
\caption{Distribution of parallax factors as a function of scan direction. The scan direction can be seen as the connection from the point (0,0) to a data point. The parallax factor (a dimensionless ratio) is indicated by the distance of the data point from the reference point (0,0). The red and blue dots refer to positive and negative parallax factors, respectively.}
\label{fig:pardir}
\end{figure}

The Hipparcos astrometric data at epoch resolution consists of one-dimensional measurements of transit times, which, by means of the reconstruction of the satellite attitude \citep{2005A&A...439..791V}, are transformed to scan phases. Any object in the Hipparcos catalogue has been scanned multiple times in different directions over the 3.3 year period of the mission. For Polaris there are 127 such measurements, of which three were rejected in the iteration for the astrometric solution. Two of these measurements were obtained during the final months of the mission, when the condition of the satellite had already significantly deteriorated. The third rejected measurement was only marginally larger than expected. The distribution over epoch of those measurement is shown in Fig.~\ref{fig:epochdistr}.
The distribution of measurements over epoch for Polaris is quite homogeneous, which is due to its position being relatively close to the ecliptic pole. The distribution of the scan directions is also homogeneous, and there is no significant correlation between the scan directions and the parallax factor (Fig.~\ref{fig:pardir}). The latter represents the effect a change in the parallax value has on a measurement. 

\begin{figure}[h]
\centering
\includegraphics[width=8cm]{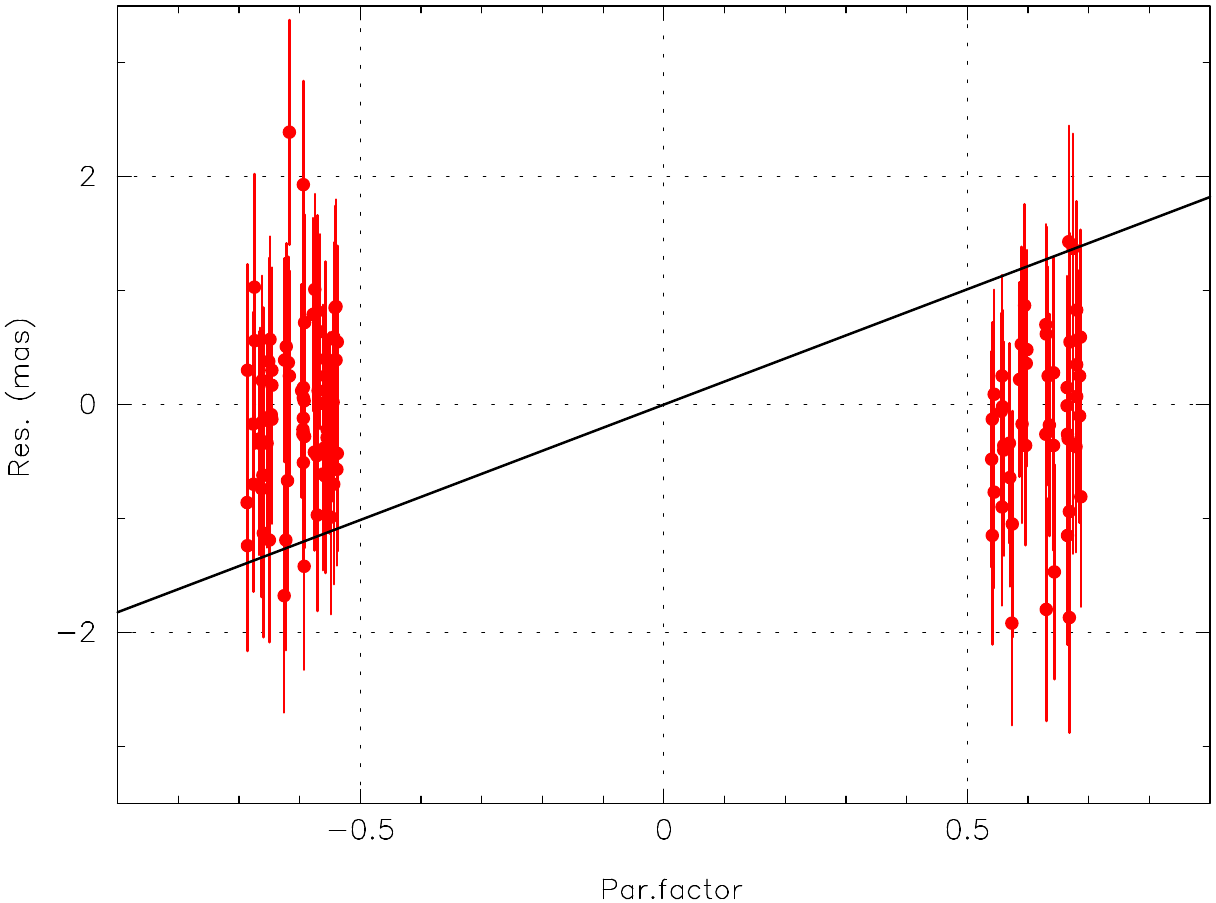}
\caption{Distribution of measurement residuals as a function of the parallax factor for the Hipparcos parallax solution of $7.54\pm0.11$~mas. Also shown, by means of the diagonal line, is the expected relation between parallax factors and residuals for the parallax of 10.1~mas that has been suggested by TKUG.}
\label{fig:parfact}
\end{figure}

The correction to the parallax is determined by the correlation between the weighted residuals and the parallax factors. For Polaris the spread in standard errors on the measurements is relatively small, ranging from 0.61 to 0.81 mas. The main reason for this is that the errors are dominated by a calibration-noise contribution from the along-scan attitude reconstruction, with a relatively small contribution from the photon-noise statistics. There are therefore no individual measurements with relatively high weight that could dominate or distort the astrometric solution. The distribution of residuals as a function of parallax factor is shown in Fig.~\ref{fig:parfact}.  
To reconcile the Hipparcos data for Polaris with a parallax of 10.1~mas it would take systematic errors in the epoch astrometric data that are only depending on the parallax factor, as the distributions over scan direction and epoch are not correlated with the parallax factor. The errors would need to be at a level of about 1.5~mas, which is more than twice the standard errors on the measurements. Considering that the goodness-of-fit of the current astrometric solution is very good, any such systematic errors would need to be very strongly correlated with the parallax factors of the individual measurements. This anomalous behaviour of the residuals then needs to be restricted to Polaris, or at most to a very small number of stars in the catalogue. If it were more general, errors at this level would show up in the distribution of negative parallaxes, which is contrary to what has been observed \citep{2007A&A...474..653V}. It seems therefore that the Hipparcos data cannot in any way support the shorter distance and larger parallax for Polaris suggested by TKUG.

\section{Other arguments used to suggest a larger parallax for Polaris}
\begin{figure}[h]
\centering
\includegraphics[width=7cm]{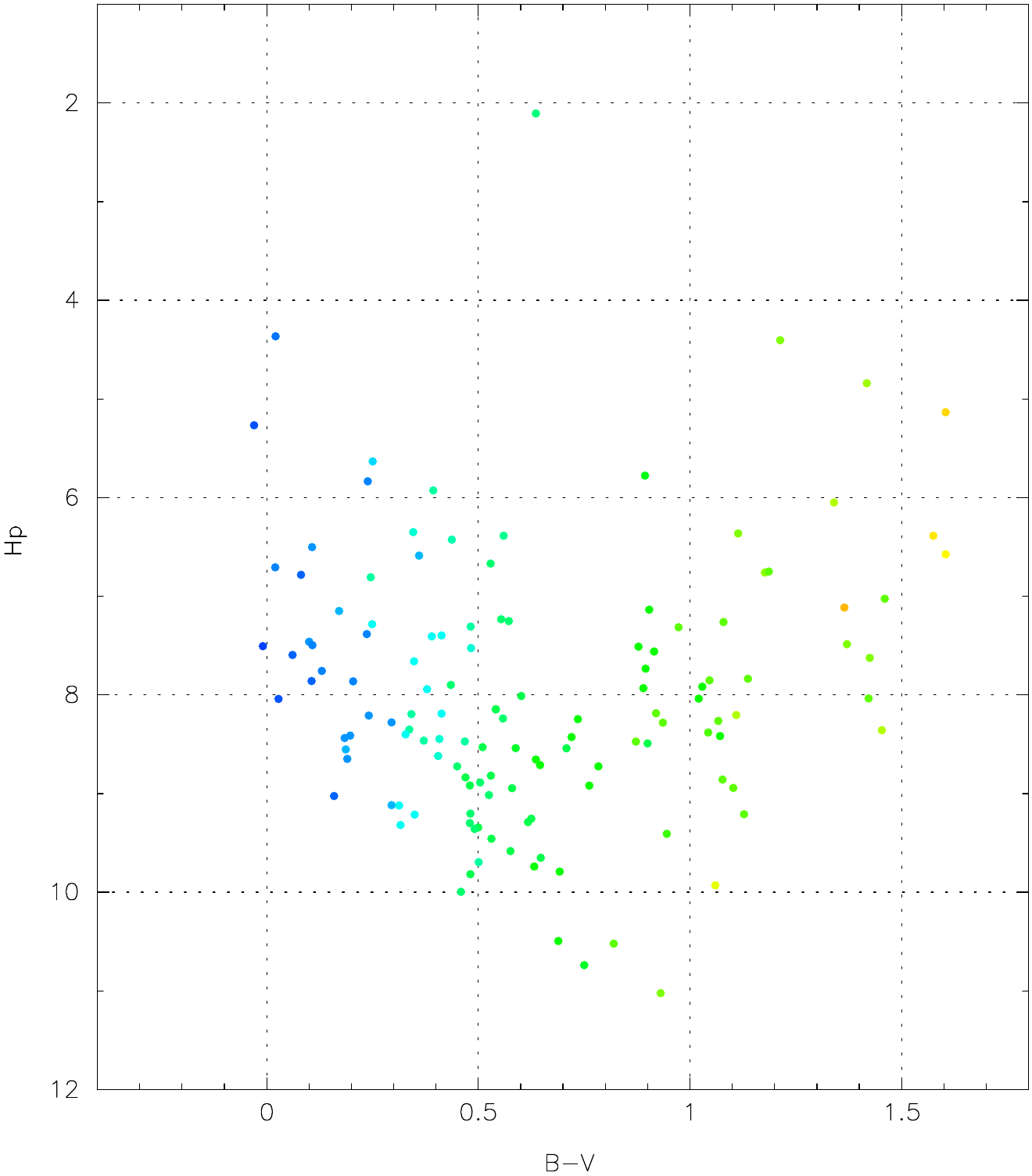}
\caption{Colour-magnitude diagram for stars within a 3 degrees radius from Polaris, and with relative parallax errors below 15 per~cent. This diagram should be compared with Fig.~4 in \citet{2009AIPC.1170...59T}. The colours of the dots reflect the spectral types of the stars, from blue (B) to red (M).}
\label{fig:hrdiagram}
\end{figure}
\begin{figure}[h]
\centering
\includegraphics[width=7cm]{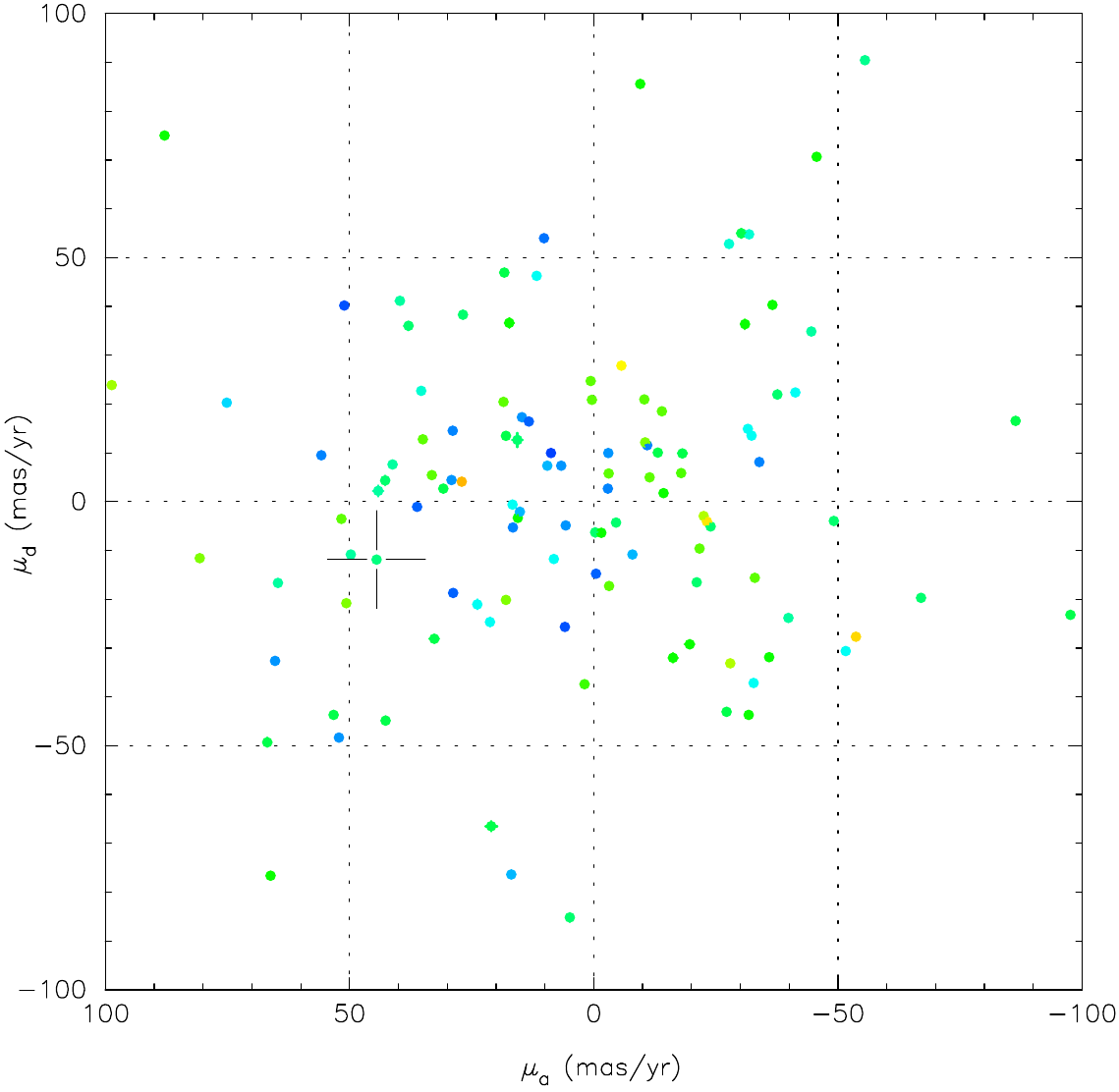}
\caption{Proper motions for 124 stars within a radius of 3 degrees from Polaris and with relative errors on the parallaxes lower than 15~per~cent. There are 18 additional stars for which the proper motions fall outside the diagram boundaries. The proper motion of Polaris is indicated by the cross-bars. Errors on the proper motions are in all cases smaller than the dot size. No clustering in proper motions is being observed. The colours of the dots reflect the spectral types of the stars, from blue (B) to red (M).}
\label{fig:propmot}
\end{figure}

One other main argument has been used in the past to suggest that the Hipparcos parallax for Polaris could be too small. Although no longer directly referred to in the paper by TKUG, it is indirectly used by referring to earlier publications on the same subject \citep{2005PASP..117..207T, 2009AIPC.1170...59T} as part of the "evidence" presented. It concerns the suggestion that there is a cluster, at a distance of 100~pc, with which Polaris is supposed to be linked. In \citet{2009AIPC.1170...59T} this cluster is presented as being observed from the HR diagram of stars within 3 degrees from Polaris based on apparent magnitudes as found in the Hipparcos catalogue. Two observations are in place here:
\begin{enumerate}
\item The diagram as presented in Fig.~4 of \citet{2009AIPC.1170...59T} as "Polaris neighbours" cannot be reproduced; in fact, the actual HR diagram for stars within 3 degrees from Polaris looks very different indeed (see Fig.~\ref{fig:hrdiagram}), and shows no sign of clustering;
\item There is no signature of a cluster in either the distribution of proper motions (see Fig.~\ref{fig:propmot}) or the distribution of parallaxes in the direction of Polaris.
\end{enumerate}
There simply is no cluster at the supposed distance of 100 pc in the direction of Polaris, and the HR diagram as produced by Turner in his 2009 paper does not appear to have any relation with reality for that field.  

\begin{figure}[h]
\centering
\includegraphics[width=7cm]{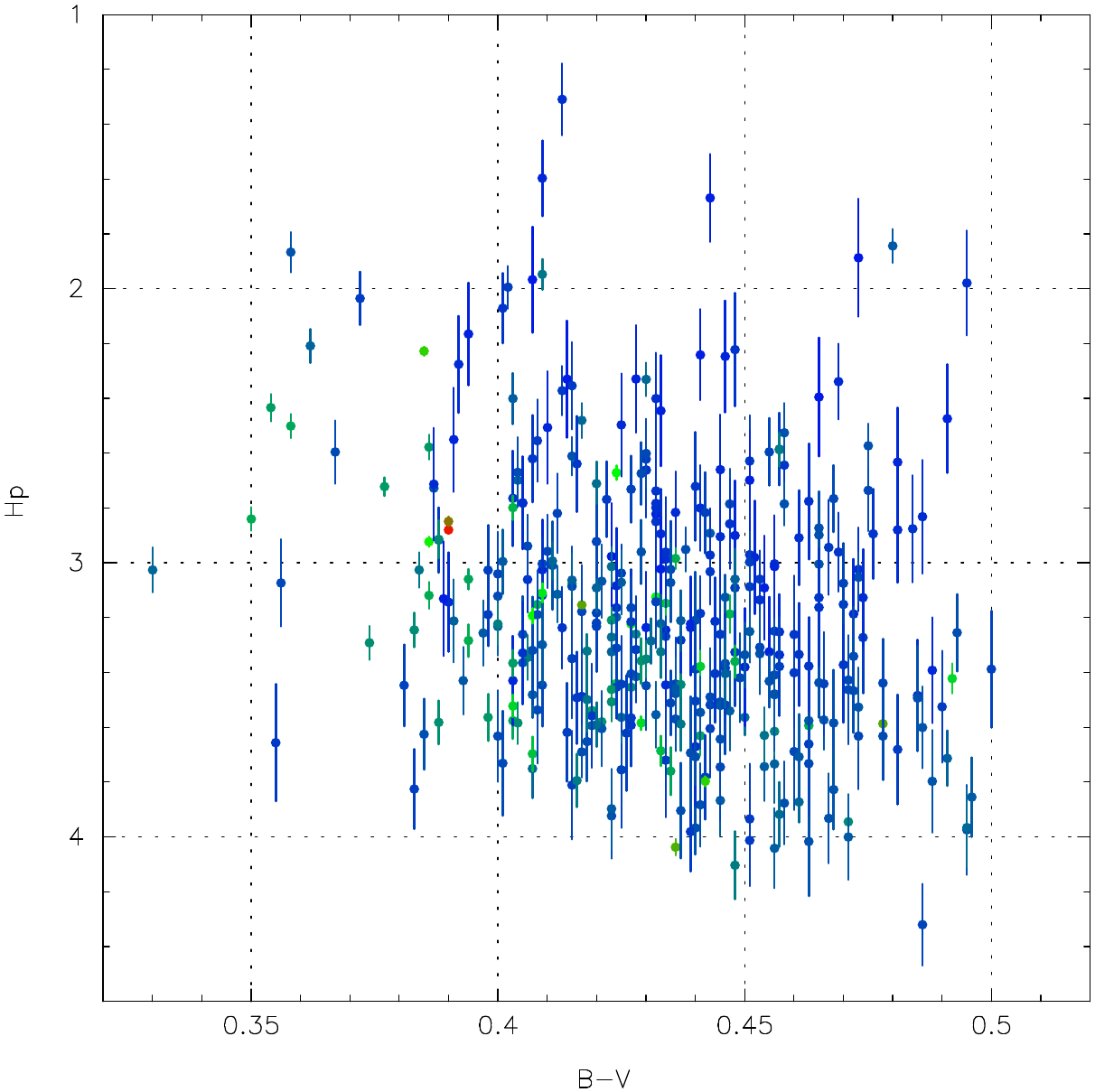}
\caption{Distribution in colours and absolute magnitudes for 389 of the 390 F3~V stars contained in the Hipparcos catalogue, and with relative errors on the parallaxes below 10~per~cent. One data point (for HIP~84034) falls outside the diagram boundaries. The error bar at each data point illustrates the uncertainty in the distance modulus. The colours of the dots reflect the parallaxes, ranging from blue (small, large distance) to red (large, short distance).}
\label{fig:f3vstars}
\end{figure}
The argument on the cluster was replaced by TKUG with the suggestion that the distance of Polaris could be determined through main-sequence fitting of the F3~V companion of Polaris, Polaris B, of magnitude V=8.60, and B$-$V=0.39 \citep{1977PASP...89..550T}. Through main-sequence fitting, TKUG derived a distance modulus of 5.0, which therefore must be based on an absolute magnitude of V=3.6. The Hipparcos catalogue contains 390 stars for which the spectral type is given as F3~V, and for which the relative error on the parallax is lower than 10~per~cent. Figure~\ref{fig:f3vstars} shows the distribution in absolute magnitude and colour for these stars. The mean absolute magnitude for this selection of stars is Hp=3.17, with a standard deviation of 0.48. The equivalent in V$_\mathrm{J}$ is $3.08\pm 0.48$ \citep{esa97}. The mean colour is 0.434 with a standard deviation of 0.029. There is a significant correlation coefficient of 0.237 between the magnitude and colour distributions, which simply reflects the spread over the B$-$V colour index among the F3~V stars. The Hipparcos measurement of the parallax implies an absolute V magnitude for Polaris~B of V=2.99. This value, as well as the value given by TKUG for Polaris~B, falls within the observed distribution of F3~V stars, with a somewhat higher probability for the Hipparcos estimate. It is clear that Polaris B can not provide conclusive evidence for either claim.

\section{Conclusions}

There may be a discrepancy in the luminosity predictions or measurements as obtained with different methods, but the uncertainty is not in the determination of the Hipparcos parallax for Polaris. There is nothing in the data from which the parallax has been derived that could possibly justify a correction by nearly 2.5~mas. There is no other direct observational support for such an adjustment either. The explanation for a discrepancy as presented by TKUG must therefore be found elsewhere, and the most obvious place to look for it is the pulsation mode for Polaris as assumed by TKUG. In particular as there does not appear to be any discrepancy on the luminosity if Polaris is not assumed to be pulsating in fundamental mode \citep{2007MNRAS.379..723V}. The observed changes in period may then, on the other hand, also become more interesting by being an unexpected behaviour. 

\bibliographystyle{aa} % style aa.bst
\bibliography{MyReferences} % 
\end{document}